\documentclass[aps,prb,preprint,graphicx]{revtex4}

\draft

\begin{document}

\title{Soliton tunneling transistor}

\author{J. H. Miller, Jr. \cite{Address}, G. C\'{a}rdenas, A. Garc\'{\i}a,
W. More, and A. W. Beckwith}
\address{Department of Physics and Texas Center for Superconductivity \\
University of Houston, Houston, Texas 77204-5506}

\author{J. P. McCarten}
\address{Eastman Kodak Company \\
1999 Lake Avenue, Rochester, NY 14650-2008}

\date{\today}

\begin{abstract}
We report on a macroscopic version of the single-electron
transistor (SET), which we call the soliton tunneling transistor
(STT). The STT consists of a gate capacitor coupled to a
NbSe$_{3}$ crystal with a charge density wave (CDW). We find that
the current-voltage characteristic of an STT is periodically
modulated by the gate voltage, as in the SET, except that the
measured periodicity corresponds to a macroscopic displacement
charge. These results appear to be consistent with time-correlated
quantum nucleation of solitons and antisolitons [see Phys. Rev.
Lett. {\bf 84}, 1555 (2000)]. We discuss how the microscopic
degrees of freedom within the condensate might enable quantum
behavior at high temperatures, and report on preliminary modeling
studies using a coupled-phase Hamiltonian to interpret our
results.
\end{abstract}


\maketitle

\section{Background and Motivation}

A wide class of nonperturbative phenomena in field theory can be
understood in terms of quantum tunneling. A well-known example is
the quantum decay of the false vacuum, \cite{Coleman} which has
been of broad scientific interest in cosmology \cite{Tryon,Linde}
and other fields \cite{Caldeira} for over two decades. In three
dimensions, the boundary between the bubble of true vacuum and the
surrounding false vacuum is a type of topological defect known as
a domain wall. A variety of topological defects in condensed
matter systems have been proposed to nucleate via quantum or
thermal fluctuations. These include vortex-antivortex pairs and
vortex rings in superconductors, \cite{Kosterlitz,Duan}
superfluids, \cite{Rayfield,Volovik} and Bose-Einstein
condensates; \cite{Anderson} dislocation pairs in Wigner crystals
\cite{Chui} and vortex lattices; \cite{Blatter} phase-slip vortex
rings in charge density waves (CDWs); \cite{Zaitsev,Matsukawa}
charge (or flux) soliton-antisoliton pairs in density waves
\cite{Maki,Bardeen,Krive,Maiti} (or Josephson junctions
\cite{Widom}); and soliton-like domain walls surrounding
cigar-shaped bubbles of true vacuum in three-dimensional CDWs.
\cite{Krive2} The quantum decay of the metastable false vacuum of
a scalar field $\phi$, accompanied by the creation of solitons and
antisolitons in (1+1) dimensions \cite{Stone,Voloshin,Kiselev} (or
soliton domain walls in (3+1) dimensions \cite{Dias}), has been
studied extensively in quantum field theory.

A CDW is a condensate \cite{Gruner} in which the electronic charge
density in a quasi-1-D metal is modulated,
$\rho(x,t)=\rho_{0}(x,t)+\rho_{1}\cos[2k_{F}x-\phi(x,t)$]. Here
$\rho_{0}(x,t)$ contains the background charge of the condensed
electrons, and an excess or deficiency of charge proportional to
$\pm \, \partial\phi/\partial x $. Weakly pinned density waves
(DWs) and weakly-coupled Josephson junctions (JJs) can be
approximately described by a phase, $\phi(x,t)$, in a sine-Gordon
(s-G) potential. They are dual in that the roles of charge and
flux are interchanged, as well as those of current and voltage.
The current in a density wave is $I=(Q_{0}/2\pi)\,\partial\phi/
\partial t$, where $Q_{0} \sim 2N_{ch}e$ and $N_{ch}$ is the
number of parallel chains, whereas the voltage across a JJ is
$V=(\Phi_{0}/2\pi)\,\partial\phi/\partial t$, where
$\Phi_{0}=h/2e$. Charge (flux) solitons in a density wave (JJ)
carry a charge (flux) of $\pm \, Q_{0}(\pm \Phi_{0})$ (see Table
I). The width of a Josephson vortex is roughly the Josephson
penetration length, $\lambda_{J} \propto J_{c}^{-1/2}$ whereas, in
a DW, the soliton width is $\lambda_{0}=c_{0}/\omega_{0}$, where
$c_{0}$ is the phason velocity and $\omega_{0}$ is the pinning
frequency. Thus, $\lambda_{0}$ will increase with decreasing
impurity concentration (as $\omega_{0}$ decreases), and may
approach the distance between contacts in extremely pure samples.
This is equivalent to approaching the short junction limit, $L <
\lambda_{J}$, in a JJ, where the {\em V-I\/} curves become
significantly less rounded.

Krive and Rozhavsky (KR) \cite{Krive} point out the existence of a
Coulomb blockade threshold, due to the electrostatic energy of the
charged solitons and antisolitons, for the creation of
soliton-antisoliton ($S-S\,'$) pairs in density waves. A more
recent paper \cite{Miller} proposes an analogy to time-correlated
single electron tunneling to explain the observed lack of DW
polarization below threshold, and to interpret key features of DW
dynamics, such as coherent oscillations and narrow-band noise,
above threshold. Observations of Aharonov-Bohm (A-B) oscillations
\cite{Latyshev} in the magneto-conductance of CDWs in NbSe$_{3}$
crystals with columnar defects provide compelling evidence for
quantum transport. Additional evidence for quantum behavior is
found in {\em rf\/} experiments \cite{Miller2,Richard,Miller3}
that show good agreement with photon-assisted tunneling theory,
and the observed Zener-like {\em I\/-V\/} characteristics.
\cite{Thorne} This Zener-like behavior is expected in the quantum
picture when $L \gg \lambda_{0}$ and the {\em I\/-V\/} curves are
rounded. The linear {\em I\/-V\/} curves seen in the extremely
pure NbSe$_{3}$ samples suggest a possible Coulomb blockade
mechanism, the dual of which is also suggested by the by the
linear {\em V\/-I\/} curves seen in cuprate JJs in the short
junction limit ($L < \lambda_{J}$).

Evidence against classical depinning includes bias-independent
{\em rf\/} and microwave responses below threshold in CDWs
\cite{Miller3,Zettl} and Wigner crystals, \cite{Li} and the
observed small phase displacements of charge \cite{Ross,Requardt}
and spin \cite{Wong} density waves below threshold. These
experiments strongly suggest that, as the electric field is
increased, $S-S\,'$ or dislocation pairs are created long before
the classical depinning field is attained. \cite{polarization}
Moreover, attempts \cite{Gammie} using a scanning tunneling
microscope (STM) to directly observe either displacement of the
CDW below threshold or sliding above threshold have been
unsuccessful. The apparent lack of sliding seen in STM
experiments, the jerky dynamics revealed by NMR experiments,
\cite{Ross} and the mode locking observed at high drift velocities
\cite{Thorne2} all suggest that the CDW spends most of its time in
the pinned state even above threshold.

The quantum interpretation of the threshold field, as a
pair-creation threshold due to Coulomb blockade, is motivated by
Coleman's paper \cite{Coleman2} on soliton pair-creation in the
massive Schwinger model. A pair of $S$ and $S\,'$ domain walls
with charges $\pm \, Q_{0}$ produce an internal field of magnitude
$E^{\ast} = Q_{0}/\varepsilon A$, as shown in Fig. 1, where $A$ is
the cross-sectional area. When a field $E$ is applied, the
difference in electrostatic energies of a state with a pair of
separation $l$ and of the ``vacuum" is $\Delta U=\frac{1}{2}\,
\varepsilon A l [(E \pm E^{\ast})^{2}-E^{2}] = Q_{0} l
[\frac{1}{2}E^{\ast} \pm E]$, which is positive when $|E| <
\frac{1}{2}\, E^{\ast}$. Conservation of energy thus forbids pair
production for fields less than a quantum threshold, $E_{T} \equiv
\frac{1}{2}\,E^{\ast}=Q_{0}/2 \varepsilon A$, which appears to be
about two orders of magnitude smaller than the classical depinning
field in NbSe$_{3}$. \cite{Miller} The observed universality
relation, $\varepsilon E_{T} \sim e N_{ch}/A$, \cite{Gruner} thus
arises quite elegantly in this model. \cite{screen}

A density wave between two contacts behaves as a capacitor with an
enormous dielectric constant (Fig. 1). The initial charging energy
is $Q^{2}/2C$, where $Q$ is the displacement charge and $C =
\varepsilon A/L$. We define $\theta \equiv 2\pi Q/Q_{0} = 2\pi
E/E^{\ast} = \pi E/E_{T}$ and note that a displacement $\phi$ near
the middle creates a non-topological kink-antikink pair with
charges $\pm (\phi/2\pi)Q_{0}$, if $\phi=0$ at the contacts. The
washboard pinning and quadratic charging energies can then be
written as: \cite{Coleman2}

\begin{equation}\label{Energy}
 U[\phi]=\int_{0}^{L}dx\,\{u_{p}\,[1-\cos\phi(x)]+
 u_{c}\,[\theta-\phi(x)]^{2}\}
\end{equation}
\ \\
where the first and second terms represent the pinning and
electrostatic charging energies, respectively, and where $u_{p}
\gg u_{c}$ for NbSe$_{3}$. \cite{Miller} If the system starts out
in its ground state, conservation of energy will prevent tunneling
when the applied field is below threshold, $\theta < \pi \; (E <
E_{T})$, as illustrated in Fig. 2. However, when $\theta$ exceeds
$\pi$ what was formerly the true vacuum becomes the unstable false
vacuum. A bubble of true vacuum, with soliton domain walls at its
surface, then nucleates and expands rapidly (Fig. 1).
\cite{bubble} After $n$ solitons of charge $Q_{0}$ (and
antisolitons of charge $-Q_{0}$) have reached the contacts, the
charging energy becomes

\begin{equation}\label{Energy-charging}
  \frac{(Q-n\,Q_{0}^2)}{2\,C}=\frac{Q_{0}^{2}}{8\pi^{2} \, C}\;(\theta-2\pi
  n)^{2}
\end{equation}
\ \\
This series of piecewise parabolas is similar to the charging
energy of a single-electron tunnel junction, except that $Q_{0}$
now represents a macroscopic charge.

A single-electron transistor (SET) \cite{Devoret} consists of a
gate capacitor $C_{g}$ coupled to an island electrode between two
small capacitance tunnel junctions in series. The gate voltage
modulates the {\em I\/-V\/} curves between the source and drain
electrodes, with a period {\em e} in displacement charge, $Q_{g} =
C_{g} V_{g}$. The displacement charges $Q_{1,2}$ across the two
tunnel junctions are related as $Q_{2} = Q_{1} + Q_{g} + q_{0}$,
where $q_{0}$ is a phenomenological offset charge induced during
cooling. \cite{Devoret} The SET is related by charge-flux duality
to the dc SQUID. The critical voltage across an SET is a periodic
function of $Q_{g}$, whereas the critical current across a SQUID
is periodically modulated (with period $\Phi_{0}$) by the flux
$\Phi$.

The model discussed above suggests that it may be possible to
demonstrate a macroscopic version of the SET, by attaching a gate
capacitor to an island electrode near the center of a quasi-1-D
crystal with a density wave. The displacement charge induced by
the gate electrode would then periodically modulate the total
critical voltage between the source and drain electrodes. Ideally,
in the absence of any shunt conductance, the periodicity of the
gate displacement charge might be expected to be $\sim Q_{0}$.
However, screening by the normal, uncondensed electrons will tend
to reduce the effectiveness of the gate which, unlike the source
and drain contacts, cannot be driven by a current source. The
displacement charges across the two segments of the crystal will
be related as $Q_{2} = Q_{1} + \beta (Q_{g} + q_{0})$, where
$\beta \sim \exp (-L_{eff}/L_{s}) \ll 1$ and $L_{s} \sim 2 \pi /
k_{s}$ is a screening length. The total charging energy of the two
segments, in this idealized model, will then be

\begin{equation}\label{Total-charging-energy}
 \frac{(Q_{1}-n_{1}Q_{0})^{2}}{2\,C_{1}} +  \frac{(Q_{2}-n_{2}Q_{0})^{2}}{2\,C_{2}} =
 \frac{Q_{0}^{2}}{8 \pi^{2}} \, \left\{\frac{(\theta_{1}-2\pi n_{1})^{2}}{2\,C_{1}} + \frac{(\theta_{2}- 2\pi n_{2})^{2}}{2\,C_{2}}\right\}
\end{equation}
\ \\
where $C_{1}$ and $C_{2}$ are the capacitances of the two segments
separated by the island electrode. The analogy to the SET suggests
that a gate voltage might modulate the {\em I\/-V\/} curves
between source and drain contacts, with a periodicity $\Delta
V_{g} \sim Q_{0} / \beta C_{g}$. The gate capacitance $C_{g}$, and
hence the attainable displacement charge $Q_{g}$, may have been
too small to observe non-monotonic behavior in previous
experiments, \cite{Adelman} in which a gate electrode was
fabricated directly on the crystal to form a MOSFET-like
structure. The soliton tunneling transistor (STT), discussed in
the next section, employs a much larger, 1-$\mu$F, gate capacitor
coupled to an NbSe$_{3}$ crystal, and exhibits non-monotonic
behavior.

\section{Experimental Measurements}

Single crystals of NbSe$_{3}$ were employed in the experiments
reported here. This material forms two independent CDWs, at
Peierls transition temperatures of 145 K and 59 K, \cite{Monceau}
respectively. The Peierls gap opens up over most of the Fermi
surface (FS) below the lower transition, but leaves a small
portion of the FS intact, so that a significant concentration
($\sim 6 \times 10^{-18}$ cm$^{-3}$) of normal, uncondensed
carriers remain down to low temperatures. The geometry used in our
experiment is illustrated in the inset to Figure 3, where the
width of the crystal is exaggerated for clarity. The NbSe$_{3}$
crystal was placed onto an alumina substrate with a series of
evaporated, 25-$\mu$m wide gold contacts. The substrate was
thermally anchored to a cold-finger in the vacuum shroud of an
open cycle helium flow cryostat and the temperature was controlled
using a Lake Shore temperature controller attached to a heater
coil wrapped around the cold-finger. A Keithley programmable {\em
dc\/} current source injected the current between two contacts,
which were bonded to the crystal near the ends using silver paint.
The ``source-to-drain" voltage was measured between two additional
gold contacts, as illustrated, and a 1-$\mu$F gate capacitor was
attached to the center gold contact using silver paint. The
spacing between contacts along the crystal was 500 $\mu$m
center-to-center, and the gate capacitor was kept inside the
cryostat.

We found that substantially smaller gate capacitors (as well as
gate capacitors with longer leads) were unable to induce a
periodic modulation of the {\em I\/-V\/} characteristic. Moreover,
{\em dc I\/-V\/} (rather than differential {\em dV/dI\/})
measurements were necessary to avoid inducing a displacement
current through the gate capacitor. A programmable voltage source
was coupled to the gate capacitor via a 10-k$\Omega$ resistor,
which limited the current flowing through the crystal during
changes in gate voltage when the gate capacitor either partially
charged or discharged. The cryostat was kept inside an
electromagnetically shielded enclosure, and $V_{sd}$ was measured
with a nanovoltmeter.

The measurements were primarily carried out at 35 K. Previous
``field effect transistor" experiments \cite{Adelman} showed the
greatest modulation at around 30 K, where the threshold field is
near its minimum. We also observed the largest modulation, using
our geometry, at comparable temperatures. We attained the best
temperature stability (better than $\pm \, 0.01$ K) when the
temperature was set to 35 K, and thus chose this temperature for
most of the measurements reported here.

Figure 3 shows several plots of CDW current as a function of
source-to-drain voltage, $I_{cdw}\; vs.\; V_{sd}$, in a NbSe$_{3}$
crystal at 35 K, for different values of gate voltage $V_{g}$. The
gate voltage is seen to modulate the threshold voltage in the {\em
I\/-V\/} curves of Fig. 3. Figure 4 displays plots of
source-to-drain voltage $V_{sd}\; vs.\; V_{g}$ for three values of
total bias current above threshold. The plots exhibit roughly
periodic behavior, similar to that observed in SETs. However, in
our system, the measured periodicity $\Delta V_{g} \sim 10$ V is
consistent with a macroscopic displacement charge $\Delta Q =
C_{g} \Delta V_{g} \sim 6 \times 10^{13}\;e$, comparable to the
charge of the conducting electrons between the contacts.

The behavior shown in Fig. 4 is quite extraordinary, and appears
to be consistent with the soliton tunneling hypothesis. The number
of parallel CDW chains, $N_{ch}$, is about $10^{8}$. Thus, one
might estimate the screening parameter as follows: $\beta \sim
Q_{0}/\Delta Q \sim 2N_{ch}e / \Delta Q \sim 3 \times 10^{-6}$.
Assuming that $\beta \sim \exp(-L_{eff}/L_{s}) \sim 3 \times
10^{-6}$, and taking $L_{eff}$ to be about the distance between
contacts ($\sim 500 \; \mu$m), yields an effective screening
length $L_{s}$ of about 40 $\mu$m. However, an alternative
interpretation for the observed periodicity might be that all of
the normal electrons between the contacts participate in screening
out the displacement charge $Q_{0}$ of the CDW. Thus, the
observation that $\Delta Q / e \sim 6 \times 10^{13}$ is roughly
the number of conducting electrons between contacts may not be a
coincidence. Further work is needed to better understand the
effects of screening by the normal, uncondensed electrons.

\section{Modeling Studies}

We believe that phase-coherent, Josephson-like tunneling of many
microscopic degrees of freedom enables quantum transport to
dominate at all temperatures below the Peierls transition
temperature, $T_{p}$. On the one hand, the observation of complete
mode locking with an $ac$ source in high quality crystals
 \cite{Thorne2} shows that the phase is coherent throughout
macroscopic regions within the crystal. On the other hand,
magneto-transport experiments on NbSe$_{3}$ crystals with columnar
defects \cite{Latyshev} demonstrate Aharonov-Bohm (AB)
oscillations with a periodicity of $h/2e$, and not $h/2Ne$ as
predicted theoretically \cite{Bogachek}, where $N$ is the number
of coupled chains. These apparently contradictory results can be
reconciled by treating a density wave as a condensate containing
many quantum degrees of freedom within a phase-coherent volume.

This suggests writing down a Hamiltonian in terms of the phases of
individual standing waves created by the dressed electrons,
coupled to the $2k_{F}$ phonons, in the CDW condensate:

\begin{eqnarray}
\hat{H} & = &\sum_{n}\hat{H}_{n} +\sum_{n,n'} U_{n,n'}[1-\cos(
\phi_{n}- \phi_{n'}) ], \label{eq5}
\end{eqnarray}

\noindent where $\phi_{n} \equiv \frac{1}{2} (\phi_{n\uparrow} +
\phi_{n\downarrow}) $ , $U_{n,n'}$ is related to the coupling
between nearby chains (or transverse wavevectors), and

\begin{eqnarray}
\hat{H}_{n}   & = & E_{F} \sum_{\sigma = \uparrow,\downarrow}
\left\{ \frac{ \Pi_{n\sigma}^{2}} {2\mu}  + E^{'}_{p}
[1-\cos(\phi_{n \sigma})]
  +
E^{'}_{c} [\phi_{n\sigma}-\theta]^{2} \right\} +
U^{'}[1-\cos(\phi_{n\uparrow}-\phi_{n\downarrow})]. \label{eq6}
\end{eqnarray}

\noindent Here $\phi_{n\uparrow} (\phi_{n\downarrow})$ is the
phase (near the mid-point between the kink and antikink) of the
standing wave created by a delocalized spin-up (spin-down) dressed
electron in the condensate, $\Pi_{n\sigma} = -i \partial/\partial
\phi_{n\sigma}$ is the momentum operator, and $E_{F}$ is the Fermi
energy. The parameter $\mu = M_{F}/m$ is the Fr\"{o}hlich mass
ratio, while $E^{'}_{p} = E_{p}/E_{F}$, $ E^{'}_{c} =
E_{c}/E_{F}$, and $U^{'} = U/E_{F}$ represent normalized pinning,
charging, and coupling energies, respectively. Note that, in a
spin density wave, the spin-up and spin-down CDWs are
out-of-phase, so the last term in Eq. (6) would be
$U^{'}[1+cos(\phi_{n\uparrow}-\phi_{n\downarrow})] $ in an SDW.
The effective charge of $2e$ observed in the AB experiments
suggests that the coupling U between the two spin components is
strong compared to the other coupling terms, i.e. $U >> U_{nn'}$.
We thus take $U^{'}$ to be nonzero in our calculations, and
consider the case where $U^{'} >> E^{'}_{p} >> E^{'}_{c} $.
Alternatively, if the phases $\phi_{n\uparrow}$ and
$\phi_{n\downarrow}$ are taken to be identical (i. e., if $U
\rightarrow \infty $), our model can be treated as one in which
pairs of chains are coupled.

In the variational approach, the energy is minimized to estimate
the ground-state wavefunctions and energies for different values
of $\theta$. Because a CDW is highly dissipative, the system will
tend to quickly relax to its lowest energy state. In the
variational method, the energy expectation value

\begin{eqnarray}
E = \langle\Psi |\hat{H}|\Psi \rangle
\end{eqnarray}

\noindent is minimized by setting $\delta E = 0 $ to estimate the
ground state energy.

The wavefunctions of the system $\chi(\phi_{n\sigma})$ are
approximated as superpositions of sharply peaked Gaussians
centered at the pinning potential minima. The state of the system,
incorporating the two spin components, is described as a trial
function with properties that depend on the parameters $b_{m}$ and
$\alpha$,

\begin{eqnarray}
\chi(\phi_{n\sigma})  &=&  \sum_{m=-N}^{N} b_{m} \exp[-\alpha (\phi_{n\sigma}-2\pi m )^2 ] \\
\nonumber \\
\Psi[\phi_{n\sigma}]  &=&  \prod_{n\sigma}\chi(\phi_{n\sigma}).
\end{eqnarray}
\ \\
The coefficients $b_{m}$ asymptotically approach {\em zero\/} as
$m$ approaches {\em infinity\/}. We set $N = 2$ in Eq. (7) to
render the problem computationally tractable. For each quantum
degree of freedom $\phi_{n\sigma} $, we choose the coefficients to
satisfy the following normalization condition:

\begin{eqnarray}
\sum_{m=-2}^2  b_{m}^2 =1.
\end{eqnarray}

We have studied the problem using of a variety of parameter values
in the Hamiltonian of Eq. (5), with qualitatively similar results.
For the results illustrated here, we take $\mu = 354$
 \cite{Gruner2}, $E_{p}/E_{F} = 10^{-5} $ , $ E_{c}/E_{F} = 10^{-6}
$, and $U/E_{F} = 5 \times 10^{-3}$. Fig. 5 shows a plot of the
minimized energy $E(\theta)$. The divergence for $ |\theta| > 4\pi
$ is an artifact of our having limiting the number of coefficients
$b_{m}$ to 5, i.e. taking $N = 2$, in Eq. (7). An exact
calculation would yield a plot in which $E$ would be a periodic
function of $\theta$. In Fig. 5, the energy is reduced slightly at
the crossing points, $\theta = n\pi$, as compared to piecewise
parabolas, but Eq. (2) provides a reasonable approximation to
$E(\theta)$:

\begin{eqnarray}
E(\theta) \sim  (\theta - 2\pi n )^{2}.
\end{eqnarray}

As in the SET, the voltage across each segment of the STT is
related to the charging energy as $V_{1,2}= dE/dQ_{1,2} \propto
dE/d\theta_{1,2} $. If we use the approximation given by Eq. (10),
this yields a sawtooth function, which can be expanded as a
Fourier series:

\begin{eqnarray}
V_{1,2}=-V_{0}\, {\rm saw}(\theta_{1,2}) = \frac{V_{0}}{\pi}
\sum^{\infty}_{n=1} \frac{ (-1)^{n} } {n} \sin(n \theta_{1,2} ).
\end{eqnarray}

\noindent Fig. 6 shows a comparison between the simple sawtooth
function and the numerical result obtained using the parameters
listed above. This model has also been used to model the dynamics
of density waves and to calculate the narrow-band noise spectra,
with excellent agreement with experiment. These results will be
reported in some detail in a separate publication.

Eq. (11) is related by charge-flux duality to the current-phase
relation of a Josephson junction. The periodic behavior of the
source-to-drain voltage vs. gate voltage of an STT can readily be
understood by exploiting the duality with a dc SQUID. When each JJ
in a dc SQUID has an ideal, sinusoidal current-phase relation and
the critical currents $I_{0}$ are identical, the total current is
$I = I_{0}[\sin\varphi_{1} + \sin\varphi_{2}]$, where
$\varphi_{2}= \varphi_{1} + 2\pi\Phi/\Phi_{0}$. This yields a
total critical current, $I{c} = 2I_{0}|\cos(2\pi \Phi/\Phi_{0})|$
, which is a periodic function of the flux $\Phi$. Similarly, the
total source-to-drain voltage of an ideal STT, $V = V_{1} +
V_{2}$, where $V_{1,2}$ are given by Eq. (11), yields a critical
voltage that is a periodic function of gate voltage $V_{g} =
Q_{g}/C_{g}$, as shown in Fig. 7. Here we note that $\theta_{2} =
\theta_{1} + \theta_{g} + \theta_{0} $, where $\theta_{g} = 2\pi
\beta Q_{g}/Q_{0}$, $ \theta_{0} = 2\pi \beta q_{0}/Q_{0}$,
$Q_{g}$ is the displacement charge across the gate capacitor,
$q_{0}$ is the offset charge, and $\beta$ is the screening
parameter due to the normal electrons. Fig. 7, although rather
simplistic, provides at least a preliminary, qualitative
interpretation of our experimental results.

\section{Conclusion}

We have carried experiments which, combined with our modeling
studies, provide further evidence for a quantum mechanism of CDW
transport. The use of wave functions, in Eq. (7) and (8), is based
on the approximation that the $\phi_{n\sigma}$'s are roughly
uniform sufficiently far from the kink and antikink. An extension
of the model would employ wave {\em functionals\/} that
incorporated spatial variations of the phases $\phi_{n\sigma}(x)$.
A soliton domain wall could then be viewed as a condensate of a
microscopic quantum solitons. The wave functional approach to
quantum field theory, together with a non perturbative treatment
of the functional Schr\"{o}dinger equation, has previously been
discussed in the context of QCD, \cite{Cornwall} scalar 2-D QED,
\cite{Kiefen} and (1+1)-D quantum gravity. \cite{Benedict}

Further refinements might include the addition of disorder to the
s-G pinning potential, which would enable a description of
metastability, hysteresis, and related phenomena. Krive and
Rozhavskii \cite{Krive3} have studied the effective Lagrangian for
macroscopic quantum tunneling of a CDW in a random medium, and
have found that disorder significantly enhances the tunneling
rate. A quantum version of a modified Fukuyama-Lee-Rice (FLR)
model, \cite{Fukuyama} which included the electrostatic charging
energy due to spatial variations of the phase, would provide a
more realistic description of many aspects of the problem.
However, we believe it is also important to incorporate the
numerous microscopic degrees of freedom when calculating the
effective action relevant for tunneling. We are thus exploring a
generalization of the tunneling Hamiltonian \cite{Bardeen2} to the
case of matrix elements between wave functionals of scalar fields
representing these degrees of freedom.

The quantum decay of the false vacuum, accompanied by the
nucleation of topological defects, is a far-reaching problem,
which could impact many areas of physics. For example, topological
defects, such as flux vortices, play an important role in the
cuprates and other type-II superconductors. \cite{Blatter}
Magnetic relaxation rates that depend weakly on temperature up to
20 K, \cite{Wen} or even decrease with temperature, \cite{Xue}
suggest that Abrikosov vortices may tunnel over a wide temperature
range. Moreover, the consistently low $I_{c}R_{n}$ products of
cuprate Josephson devices suggest that Josephson vortex-antivortex
pair creation \cite{Widom} may occur when the current is much
smaller than the``classical" critical current $I_{0} \sim
\Delta/R_{N}e$. In cosmology, the existence of many matter fields
may facilitate quantum nucleation of a universe even when the
total action is large, as suggested by Hawking et al.
\cite{Hawking} Finally, the extraordinary rapidity of first-order
phase transitions, such as the palpably visible nucleation of ice
in supercooled water, suggest a possible similarity to the decay
of the false vacuum.

The authors gratefully acknowledge the valuable contributions of
Lei-Ming Xie, James Claycomb, and Carlos Ord\'{o}\~{n}ez. This
work was supported, in part, by the State of Texas through the
Texas Center for Superconductivity and the Texas Higher Education
Coordinating Board Advanced Research Program (ARP), and by the
Robert A. Welch Foundation (E- 1221). AGP and WM gratefully
acknowledge the World Laboratory Center for Pan- American
Collaboration in Science and Technology for generous support.

\newpage

\begin{figure}[h!]
  \centering
  \caption{(a) CDW phase vs. position, illustrating the production of a soliton-antisoliton domain
           wall pair. (b) Model of a CDW capacitor, showing the nucleated domain walls moving towards
           the contacts. The electric field between the domain walls is reduced by the internal
           field $E^{\ast}$.  The distances $l$, $\lambda_{0}$, and the crystal thickness are greatly exaggerated for
           clarity.}\label{}
\end{figure}

\begin{figure}[h!]
  \centering
  \caption{Potential energy vs. $\phi$ for two different values of $\theta$. Tunneling is prevented
           by conservation of energy when $\theta < \pi$.  The small energy barrier for each
           microscopic degree of freedom within the condensate enables tunneling when $\theta > \pi$,
           while the Peierls condensation energy prevents thermal hopping.}\label{}
\end{figure}

\begin{figure}[h!]
  \centering
  \caption{CDW current vs. source-to-drain voltage for several values of gate voltage at 35 K.
           (The shunt current of the normal electrons has been subtracted for clarity.)}\label{}
\end{figure}

\begin{figure}[h!]
  \centering
  \caption{$V_{sd}$ vs. $V_{g}$ for fixed values of total bias current at 35 K.}\label{}
\end{figure}

\begin{figure}[h!]
  \centering
  \caption{Computed ground state energy $E$ vs. $\theta$, showing the periodic, piecewise parabolic
           form. The apparent divergence for $\theta > 4\pi$ is an artifact of the finite number of
           coefficients used in the trial wave function.}\label{}
\end{figure}

\begin{figure}[h!]
  \centering
  \caption{Voltage as a function of $\theta$, similar to the dual of the Josephson current-phase
           relation. The sawtooth expansion with $10^{4}$ Fourier components (dashed line) is compared
           to the numerical results obtained by minimizing the energy using the trial function
           discussed in the text.}\label{}
\end{figure}

\begin{figure}[h!]
  \centering
  \caption{Predicted critical voltage vs. normalized gate voltage $\theta_{g}$ for several values of
           $\theta_{0}$ using the idealized model discussed in Section III, showing the periodic
           behavior.}\label{}
\end{figure}

\begin{table}[h]
  \centering
  \caption{Charge-flux duality between density waves and Josephson junctions.}\label{Tab1}
  \ \\
  \begin{tabular}{ccc}
    \hline\hline
     & Density Wave & Josephson junction \\
    \hline
   Soliton or antisoliton & kink w/ charge $\pm Q_{0}$ & Josephson vortex w/ flux $\pm \Phi_{0}$ \\
   Type of threshold & Threshold field $E_{T}$ & Threshold current $I_{T}$ \\
   Transport characteristic & $\;I \,vs. \,V, \; I=2\pi Q_{0} \partial \phi /\partial t\;$ & $\;V \,vs. \,I, \; V=2\pi \Phi_{0} \partial \phi /\partial t\;$ \\
   \hline\hline
  \end{tabular}
\end{table}


\begin{references}
\bibitem{screen} Screening by the normal carriers may enhance $\varepsilon$. If $\varepsilon$ scales as
$n_{\rm n}/n_{\rm i}^{2}$, where $n_{\rm i}$ is the concentration
of pinning impurities, then $E_{T}$ will vary as $n_{\rm i
}^{2}/n_{\rm n}$.
\bibitem{bubble} Actually, many small ``bubbles" will nucleate, followed by annilihation of colliding
$S$ and $S\,'$ domain walls, which causes the phase to advance by
$2\pi$ between the contacts.
\end{references}
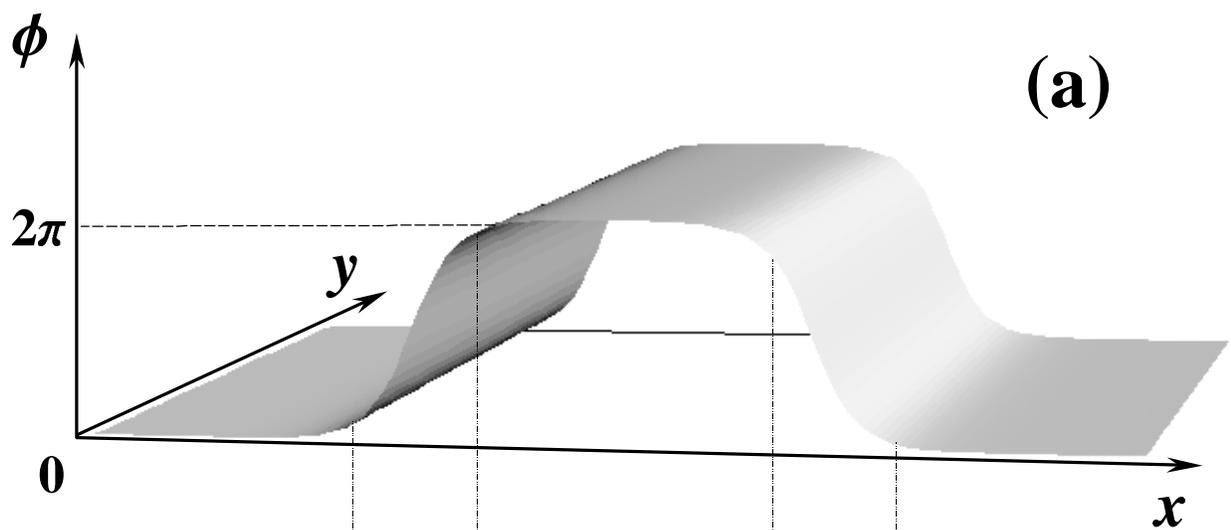
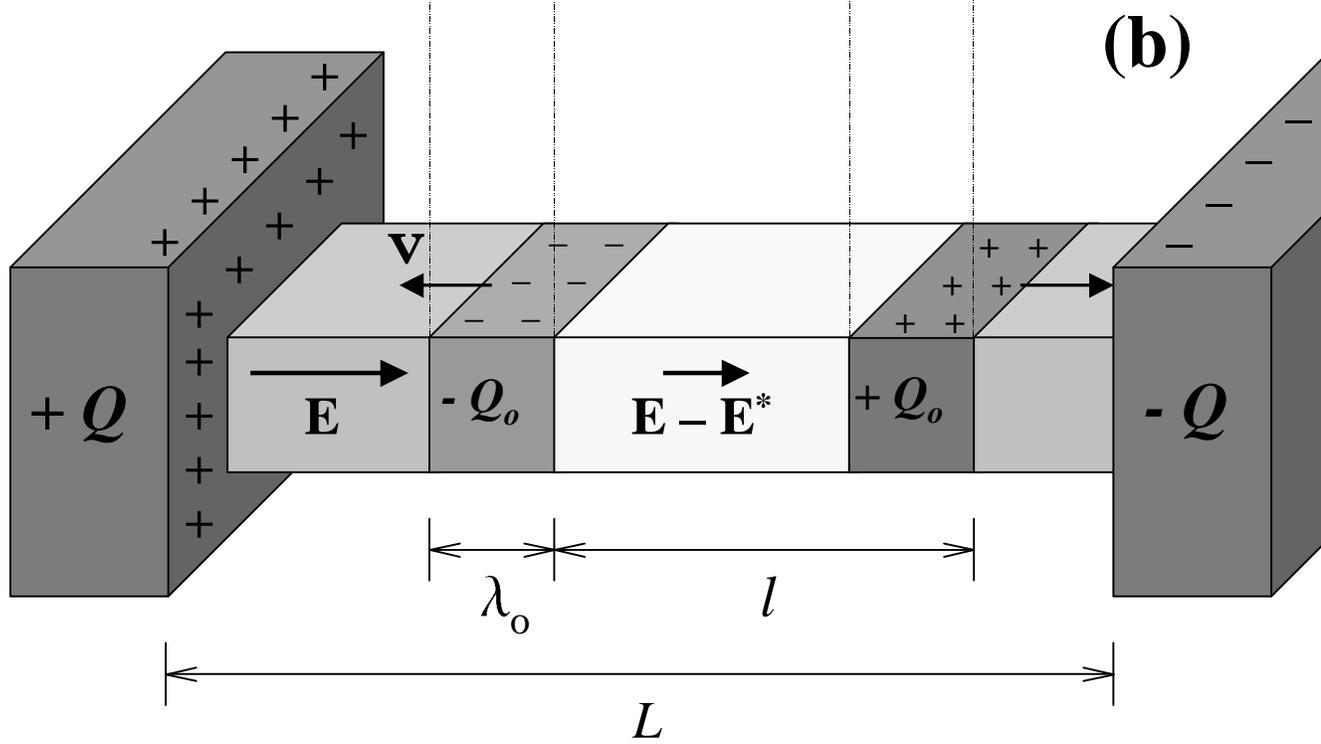
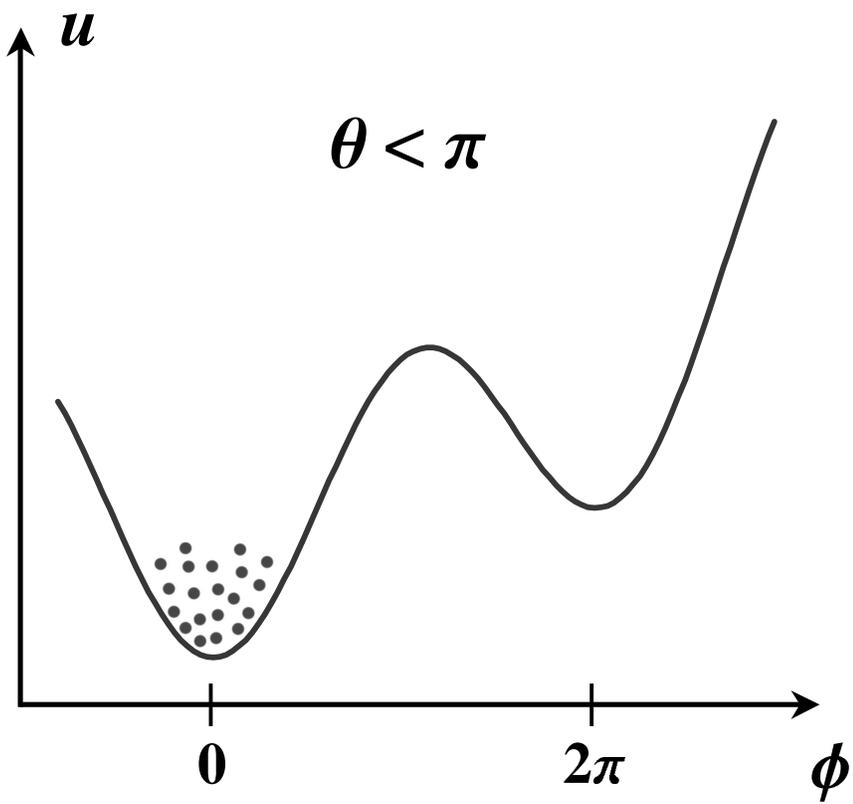
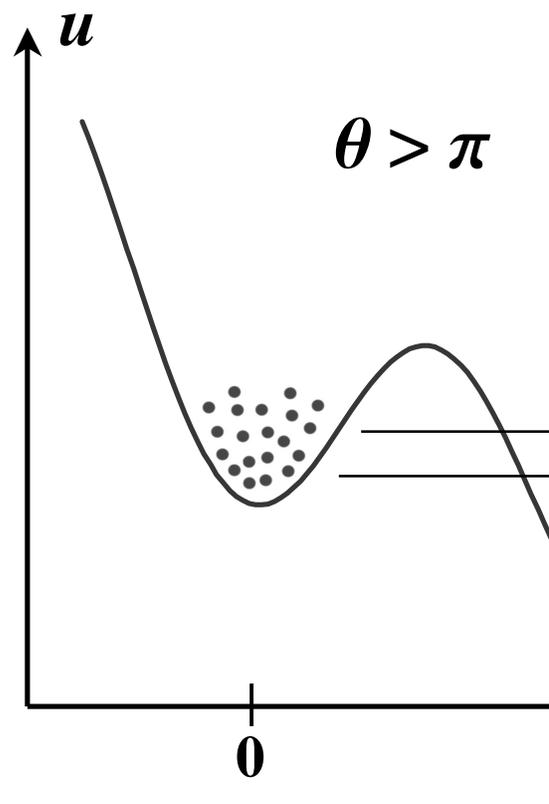
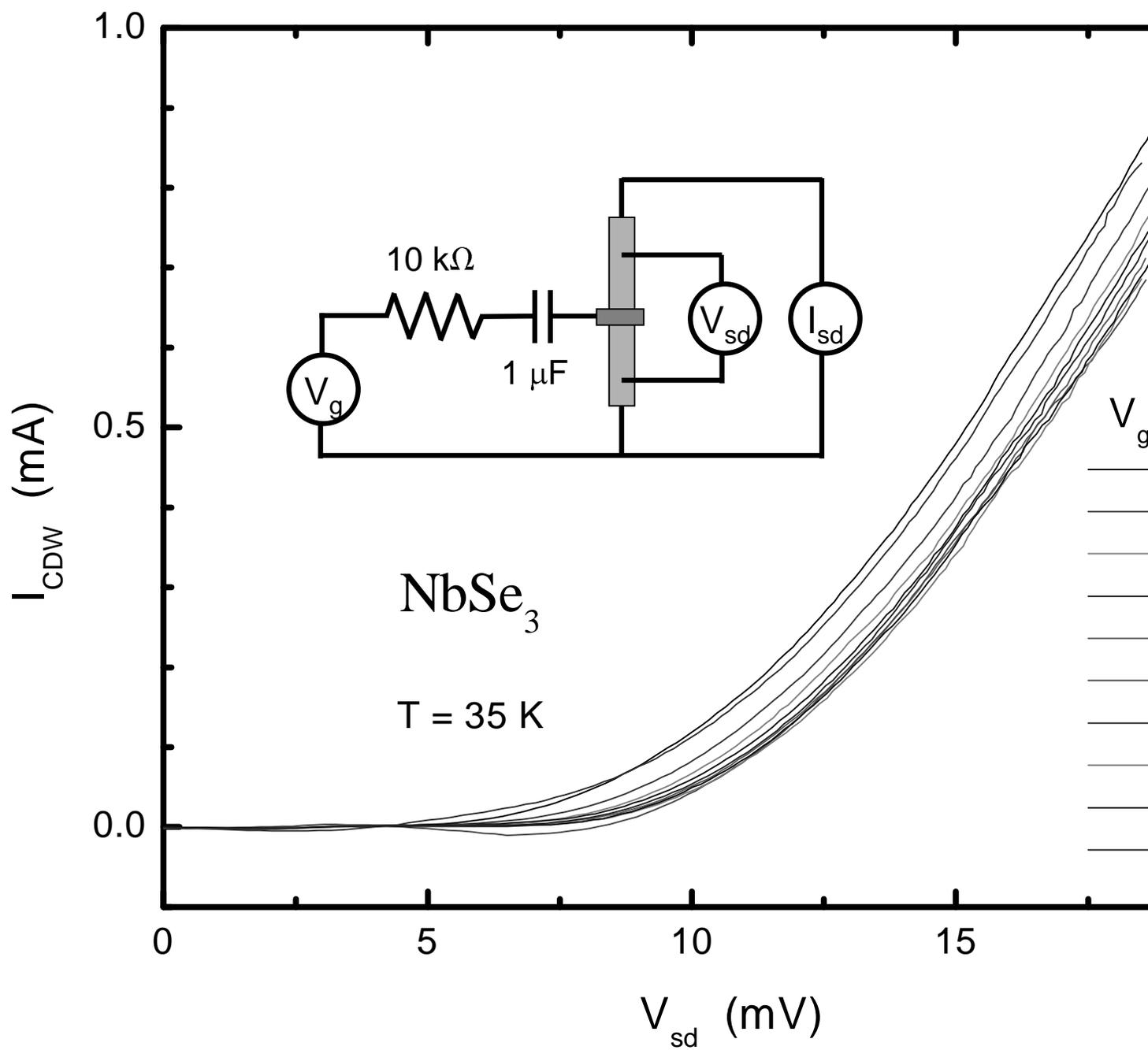
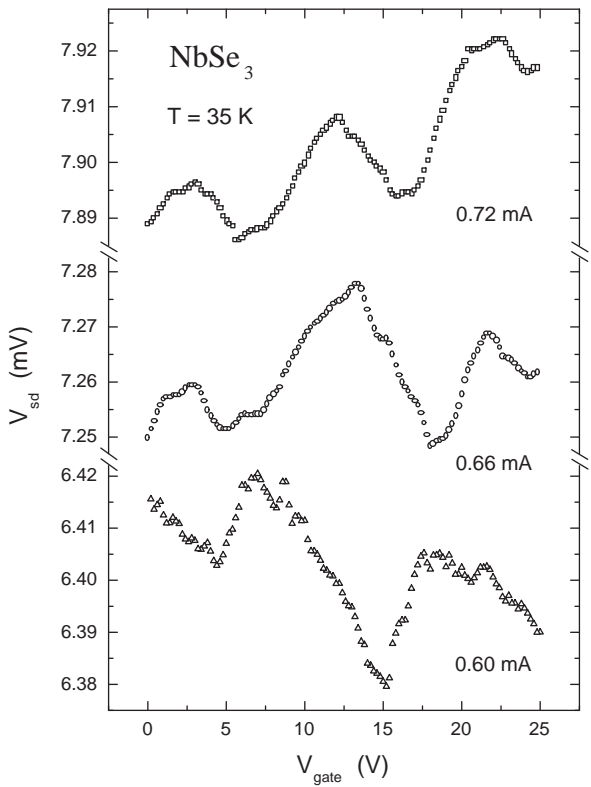
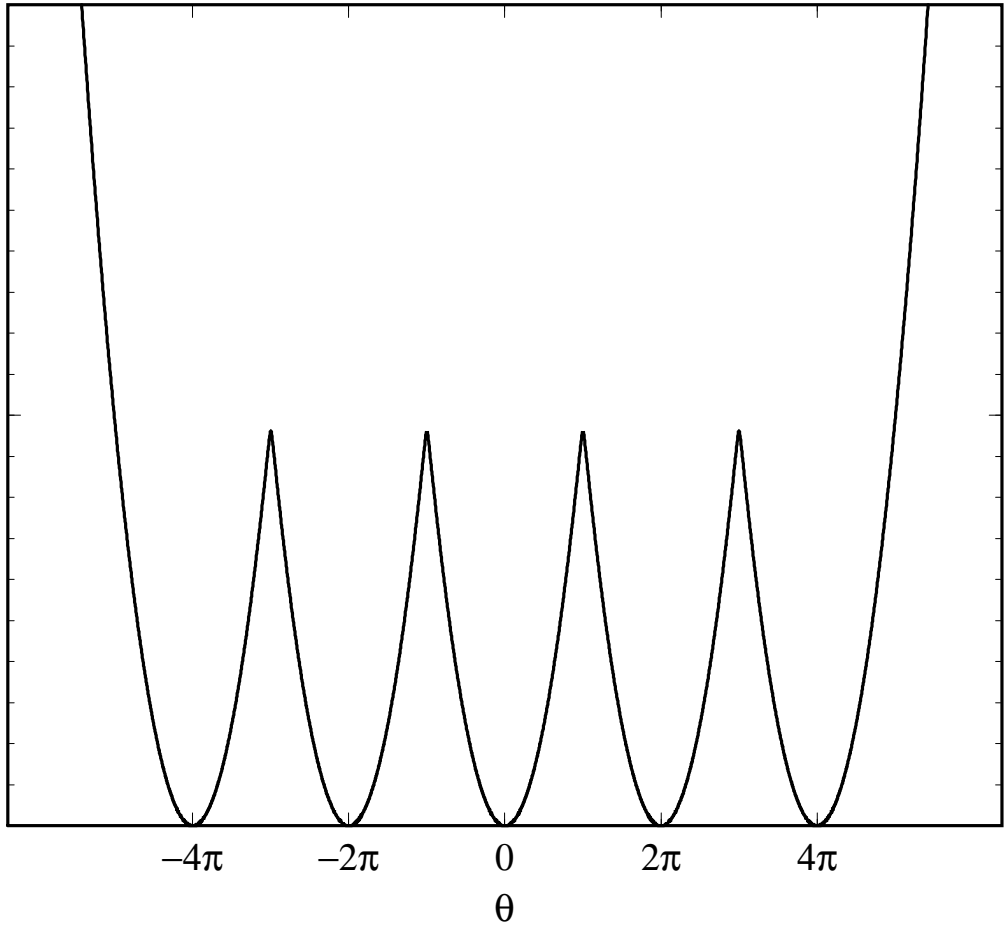
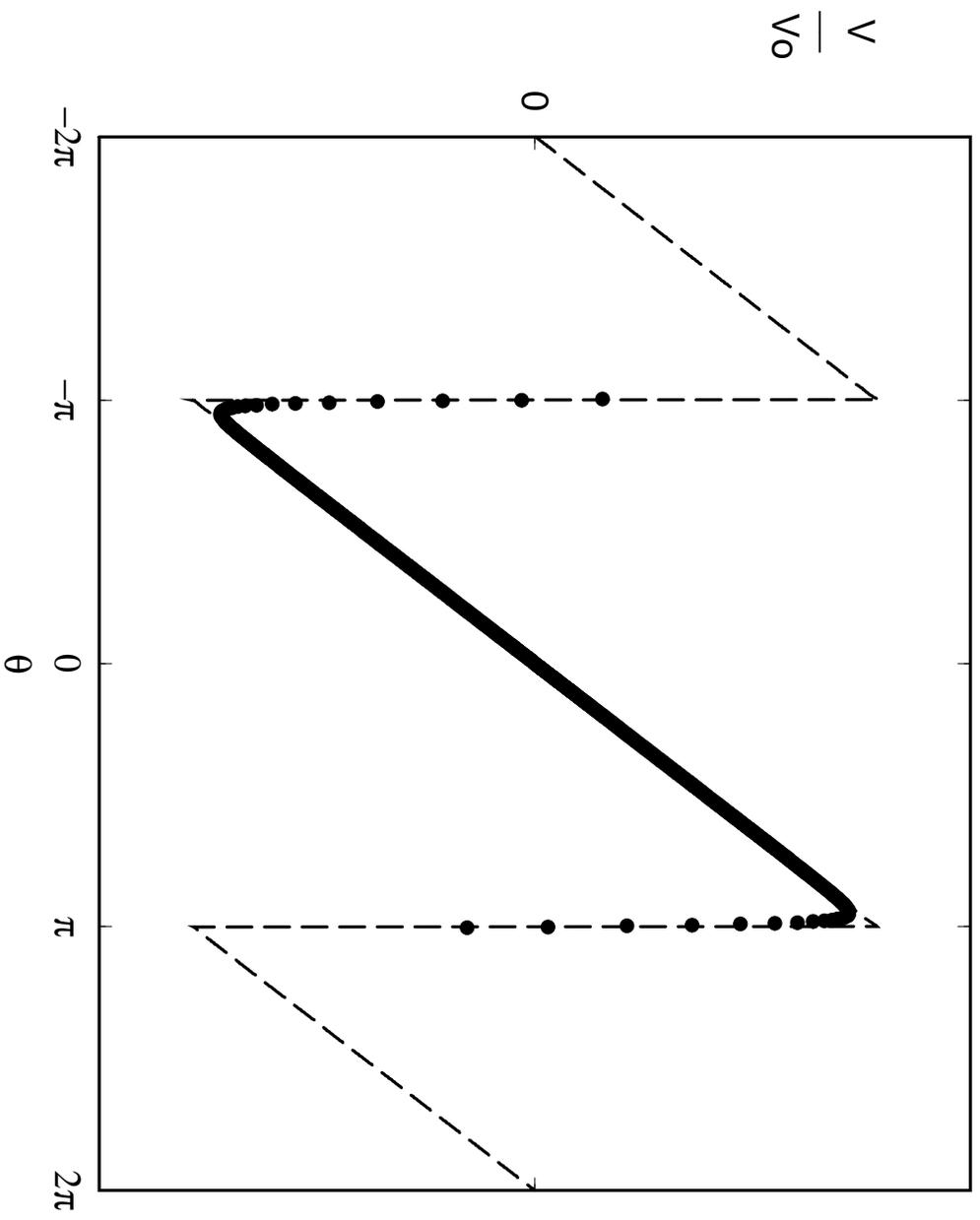
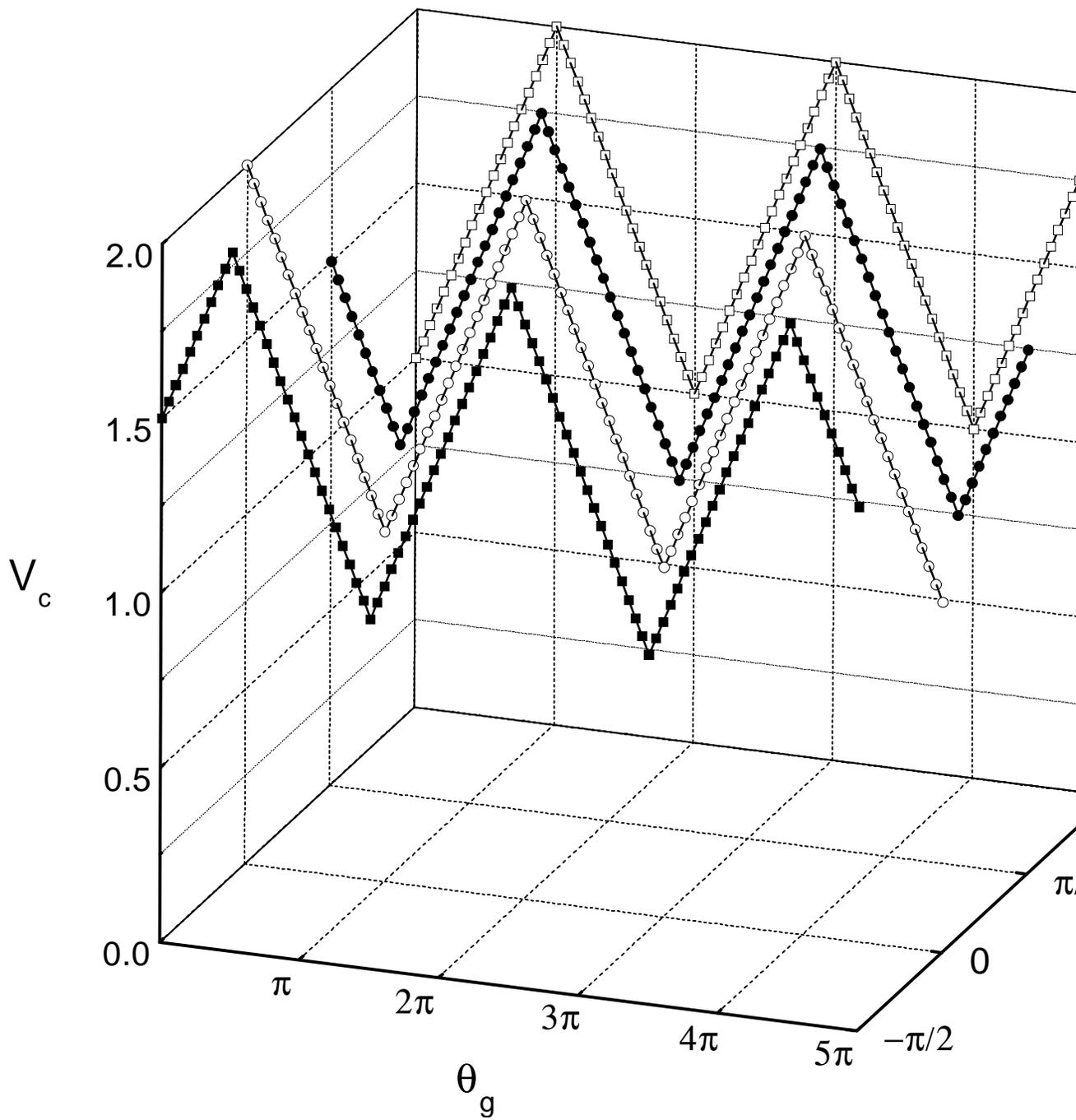
\end{document}